\documentstyle[12pt]{article}

\newcommand{\la}[1]{\label{#1}}

\newcommand{\pspict}[2]{
  \vspace*{#1}
  \special{dvitops: import #2 \the\textwidth #1}
}

\newlength{\numpit}

\newlength{\indeksinpituus}

\newlength{\ypit}

\newcommand{\be}{\begin{equation}}
\newcommand{\ee}{\end{equation}}
\newcommand{\ba}{\begin{eqnarray}}
\newcommand{\ea}{\end{eqnarray}}

\newcommand{\eq}{eq.~}

\newcommand{\fig}{fig.~}

\newcommand{\dd}{{\rm d}}

\newcommand{\lbar}{\bar{\lambda}}
\newcommand{\sigmainfty}{\sigma_{{\rm p}}}
\newcommand{\Rcr}{R_{{\rm cr}}}
\newcommand{\RLcr}{ R_{{\rm Lcr}} }
\newcommand{\RTcr}{ R_{{\rm Tcr}} }

\title{
{\bf BUBBLE FREE ENERGY\\
IN COSMOLOGICAL\\
 PHASE TRANSITIONS}
}
\author{ {\bf J. Ignatius}\thanks{
Bitnet: {\em IGNATIUS@FINUHCB}  }
\\
{\em  Department of Theoretical Physics}
\\
{\em  P.O. Box 9 (Siltavuorenpenger 20 C)}
\\
{\em  SF-00014 University of Helsinki, Finland} \\
}

\date{January 7, 1993}

\begin{document}
\maketitle

\vspace*{-12cm}
\hfill Preprint HU-TFT-92-52
\vspace*{12cm}

\begin{abstract}
Free energy as a function of temperature and the bubble radius is determined
for spherical bubbles created in cosmological first order phase transitions.
The phase transition is assumed to be driven by
an order parameter (e.g. a Higgs field) with quartic potential.
The definition of the bubble radius and the corresponding generalized,
curvature-dependent surface tensions are discussed.
In the free energy expansion in powers of the inverse radius, the coefficients
of the curvature term and the constant term are also calculated.
\end{abstract}

\newpage

Recently the detailed mechanism of the electroweak phase transition,
which depends on the free energy of the nucleating bubbles, has gained much
interest~\cite{EW,EIKR,Dineetal},
mainly because it is believed that this transition
has had a significant effect on the baryon asymmetry of the Universe.
The purpose of this Letter is to study the free energy of bubbles
which are created in cosmological first order phase transitions.
In particular, we study its expansion in $1/R$, the definition of $R$
(the bubble radius)
and an $R$--dependent surface tension.
The results can also be used to discuss
the observation~\cite{Mardor&Svetitsky} that in the MIT bag model
small hadron bubbles exist already above the quark--hadron phase
transition temperature.

The phase transition of the electroweak theory, and of the possible
grand unified theory as well, is generally thought to be driven by
a Higgs field. The theory could contain several Higgs fields, but one
combination can be used as an order parameter. For QCD, the choice of
the order parameter is not so clear, since the theory has no classical
potential. For example, the energy density
could act as an effective order parameter field~\cite{Csernai&Kapusta}.
It should be noted here, that even though we consider only first order phase
transitions in this Letter, in reality the order of the electroweak or
the quark--hadron phase transition is not known.

We begin by writing
the action for the bosonic order parameter field in the high temperature
approximation:
\be
  S = \frac{1}{T} \int \dd ^3 x \left[ \frac{1}{2} ( \nabla \phi )^2
                        + V( \phi, T )  \right] \; .    \la{S}
\ee
The effective potential, which is commonly used to describe a first order
electroweak phase transition is~\cite{Linde83,EW,EIKR,Dineetal}
\be
  V(\phi,T) = \frac{1}{2} \gamma (T^2 - T_0^2) \phi^2
             - \frac{1}{3} \alpha T \phi^3
             + \frac{1}{4} \lambda \phi^4 \; .
  \la{V}
\ee
The potential for the energy density, which was used
in~\cite{Csernai&Kapusta} for QCD, is also quartic.
This potential $V(\phi,T)$ in \eq(\ref{V}) should be regarded as a
phenomenological expansion for the effective order parameter,
valid in the vicinity of $T_c$.

First, we will describe the relevant thermodynamical properties of
the potential in \eq(\ref{V}), based on~\cite{EIKR}.
In the thermodynamical limit, the transition occurs when the heights of
the two minima are equal. This happens at the temperature $T_c$,
\be
  T_c = \frac{T_0}{\sqrt{ 1-\frac{2}{9}\frac{\alpha^2}{\lambda \gamma} }} \; .
\ee
In cosmology, the transition takes place after some supercooling at
a temperature lower than $T_c$.
It is convenient to express the temperature dependence of several
quantities by using the function $\lbar(T)$,
\be
  \lbar (T) =  \frac{9}{2} \frac{\lambda \gamma}{\alpha^2}
                 \left( 1 - \frac{T_0^2}{T^2} \right) \; ,    \la{lbar}
\ee
which satisfies $\lbar (T_0)=0$, $\lbar(T_c) = 1$.

The pressure in the low temperature (broken symmetry) phase minus the pressure
in the high temperature (symmetric) phase is
\be
  \Delta p(T) = \frac{\alpha^4}{24 \lambda^3} T^4
   \left\{  \frac{8}{27} \lbar(T)^2 - \frac{4}{3} \lbar(T) + 1
           + [ 1 - \frac{8}{9} \lbar(T) ]^{3/2}  \right\} \; .
\ee
This pressure difference is positive for temperatures smaller than $T_c$.
At $T_c$, the correlation lengths in both phases are equal to
\be
  l_c = \frac{ 3 \sqrt{\lambda} }{\sqrt{2} \alpha} \frac{1}{T_c} \; .
\ee
The action in \eq(\ref{S}) determines two additional thermodynamical
quantities which are useful for us; namely the latent heat
\be
  L = \frac{4}{9} \frac{\alpha^2 \gamma}{\lambda^2} T_0^2 T_c^2    \la{L}
\ee
and the thermodynamical or planar surface tension
\be
  \sigmainfty = \frac{2\sqrt{2}}{81}
             \frac{\alpha^3}{\lambda^{5/2}} T_c^3 \; .
\ee
The latter quantity is denoted by $\sigmainfty$,
since later on we will define a generalized surface tension $\sigma(R)$.

The latent heat coming from the decrease of the effective
relativistic degrees of freedom is included in \eq(\ref{L}).
The constants $T_0$, $\gamma$, $\alpha$ and $\lambda$
are to be chosen so that the potential quantitatively correctly describes
the phase transition. These four constants can be expressed
in terms of the more physical quantities $T_c$, $l_c$, $L$ and
$\sigmainfty$~\cite{Kajantie92}:
\be
\begin{array}{rclcrcl}
  T_0  &   = & \mbox{\Large{$ \frac{T_c}
                  { \sqrt{ 1+\frac{6\sigmainfty}{L l_c} } } $} } \; , & &
  \gamma & = & \mbox{\Large{$ \frac{L + 6\sigmainfty/l_c}
                  {6\sigmainfty l_c T_c^2} $} }\; ,  \\
  \alpha & = & \mbox{\Large{$ \frac{ \sqrt{3} }
                        { \sqrt{2 \sigmainfty} l_c^{5/2} T_c } $} } \; , & &
  \lambda &= & \mbox{\Large{$ \frac{1}{3 \sigmainfty l_c^3} $} } \; .
\end{array}
\ee

Values of $T_0 / T_c$ which are smaller than
$\sqrt{5}/3  \approx  0.75$ are not very natural~\cite{EIKR}.
However, when the potential is required to be
valid only in the vicinity of $T_c$, also smaller values are possible.
Then, if one requires that the vacuum expectation value of
the low temperature phase increases with decreasing temperature near $T_c$,
it follows that $T_0 / T_c > 1 / \sqrt{3}  \approx  0.58$ .

The decay rate of the metastable vacuum can be calculated using standard
methods~\cite{Coleman77,Linde77&81,Linde83}.
Probability of tunneling per unit time per unit volume in the high
temperature approximation is
\be
  P(T) = a(T) \: e^{ - S_{\rm cr} (T) } \; ,
\ee
where
the pre-exponential factor $a(T)$ is expected to be of the order of $T_c^4$.
The critical action $S_{\rm cr} (T)$ is the value of the action (\ref{S}),
with the appropriate boundary conditions, in the $O(3)$--symmetric extremum.
More physically speaking, it is the free energy $F_{\rm cr} (T)$,
which is needed to form a critical bubble, divided by the temperature.

The critical action can be analytically approximated in the small
and large relative supercooling regimes~\cite{Linde83},
and in the full range it has been calculated
numerically~\cite{EIKR,Dineetal}.
In terms of the function $\lbar (T)$ from \eq(\ref{lbar}),
\be
  S_{\rm cr} (T) = \frac{ F_{\rm cr} (T) }{ T }
  = \frac{2^{9/2}}{3^5 \pi} \frac{\alpha}{\lambda^{3/2}}
    \frac{ f(\lbar) }{ (1-\lbar)^2 } \; .   \la{fdef}
\ee
A fit to the function  $f(\lbar)$ with an
\mbox{$ -0.1 $} \ldots\ \mbox{$ +1.6 \% $}
error was presented in~\cite{Dineetal}. Here we give for $f(\lbar)$
a four-parameter fit with an accuracy of $\pm 0.1 \%$~:
\begin{eqnarray}
  f(\lbar) & = & \lbar^{3/2} \left( a_0 + a_1 \lbar + a_2 \lbar^2
                + a_3 \lbar^3 + a_4 \lbar^4  \right)
    \la{fnum} \; ;  \\  & &
 \begin{array}{lrlrl}
   a_0 = & 15.63628, & a_1 = & -18.03398, & a_2 = 2.39731,   \\
   a_3 = & - 0.86504, & a_4 = & 1.86543.
 \end{array} \nonumber
\end{eqnarray}
This fit is exact in the small relative supercooling ($\lbar=1$) limit.
The function $f(\lbar)$ is shown in \fig\ref{fig:f},
together with $f(\lbar) / \lbar^{3/2}$.
It is interesting to note that the latter one,
which is relevant in the large relative supercooling ($\lbar=0$)
limit~\cite{EIKR}, is an almost straight line over the full range.

Now we will turn to the main issue of this Letter:
study of the Helmholtz free energy of a spherical bubble of the
low temperature phase nucleating in the high temperature phase, and
a discussion of what effects differing definitions of the bubble radius have.
In the theory of fluid surfaces, the dependence of the surface tension on the
definition of the bubble radius is a subtle issue~\cite{Navascuesetc}.
With the exact calculation of the extremal configuration
from \eq(\ref{S}) the effect of changing the definition can precisely be
evaluated. One should bear in mind, however, that when discussing bubble
formation only the free energy has real physical importance;
surface tension as well as curvature coefficient are only auxiliary quantities.

Expanded in powers of $1/R$, i.e. in the large $R$ limit,
the free energy can be written as
\be
  F(R,T) = - \frac{4 \pi}{3} \Delta p(T) R^3 + 4 \pi \sigma_f (T) R^2
           + 8 \pi \gamma_f (T) R + 16 \pi \delta_f (T) + \ldots  \; ,
\ee
where $R$ is the bubble radius according to some specific definition.
Note that this general bubble free energy is a function of
both $R$ and $T$, whereas the free energy of a critical bubble,
$F_{\rm cr} (T)$, depends on one variable only. They are related by
\be
  F_{{\rm cr}} (T) = F( \Rcr (T), T) \; ,
\ee
where $\Rcr (T)$ is radius of the critical bubble.

It is expected that near $T_c$ the relative change of the functions
$\sigma_f (T)$, $\gamma_f (T)$, $\delta_f (T)$ is slow,
in contrast to the pressure difference $\Delta p(T)$,
since properties of the interface
do not change dramatically with temperature.
In cosmology, where we usually are discussing bubble formation only near
$T_c$, we can treat these functions as constants.
This assumption enables us to solve the general free energy $F(R,T)$ from
the critical free energy $F_{\rm cr} (T)$ and the pressure
difference $\Delta p(T)$.
Now the free energy is for large $R$, in the vicinity of $T_c$
\begin{eqnarray}
  F(R,T) & = &  - \frac{4 \pi}{3} \Delta p(T) R^3 + 4 \pi \sigmainfty R^2
           + 8 \pi \gamma R + 16 \pi \delta + \ldots   \la{F} \\
  & \equiv &  - \frac{4 \pi}{3} \Delta p(T) R^3 + 4 \pi \sigma(R) R^2 \; ,
   \la{F2}
\end{eqnarray}
where the latter form defines the generalized surface tension
(see also~\cite{KPR})
\be
  \sigma(R) = \frac{1}{4\pi R^2} \left[ F(R,T)
              + \frac{4\pi}{3} \Delta p(T) R^3 \right]  \; . \la{sdef}
\ee
The critical free energy $F_{{\rm cr}} (T)$ determines $\sigma(\Rcr (T))$,
assuming that $\Delta p(T)$ is known.
If $\Rcr (T)$ is known as well, we can solve for $\sigma(R)$.
And from $\sigma(R)$, the coefficients
in \eq(\ref{F}) can be solved:
\begin{eqnarray}
  \sigmainfty & = & \left. \sigma(R) \right| _{\frac{1}{R}=0} \; , \nonumber \\
  \gamma & = & \left. \frac{1}{2} \frac{\dd \sigma(R)}{ \dd (\frac{1}{R}) }
               \right| _{\frac{1}{R}=0}  \; , \la{gddef} \\
  \delta & = & \left. \frac{1}{4} \frac{1}{2!}
             \frac{\dd ^2 \sigma(R)}{ \dd (\frac{1}{R})^2 }
             \right| _{\frac{1}{R} =0}  \; . \nonumber
    \end{eqnarray}

We employ two different definitions for the bubble radius. The first one
is Laplace radius $R_{\rm L}$ defined by using Laplace's relation
\be
  \RLcr (T) = \frac{2 \sigma_{{\rm L}} ( \RLcr (T) )}{ \Delta p(T)} \; .
\ee
The second one is the tension radius $R_{\rm T}$, defined as the distance
at which the gradient density term of the critical action reaches its maximum.
The definition of $R_{{\rm T}}$ is illustrated in \fig\ref{fig:s3dens}.

{}From Laplace's relation, it follows that the radius of the critical bubble
and
the generalized surface tension in terms of the critical free energy are
\begin{eqnarray}
  \RLcr (T) & = & \left[ \frac{3}{2 \pi} \frac{F \left( \RLcr (T), T \right)}
                {\Delta p(T)} \right] ^{\frac{1}{3}} \; , \\
  \sigma _{{\rm L}} (\RLcr (T)) & = & \left[ \frac{3}{16 \pi}
                F \left( \RLcr (T), T \right) \, \Delta p(T)^2 \right]
                  ^{\frac{1}{3}} \; .
\end{eqnarray}
In the maximum tension definition, the radius of the critical bubble,
$\RTcr (T)$, is obtained from the numerical solution of the extremum
differential equation for the action in \eq(\ref{S})
for different values of $\lbar$.
The curvature-dependent surface tension
$\sigma _{{\rm T}} (\RTcr (T))$ is then calculated directly from
\eq(\ref{sdef}).

The two definitions for the bubble radius are compared in \fig\ref{fig:rofl}.
One notes that always $\RTcr \geq \RLcr$.
In the small relative supercooling limit, the two definitions coincide.
On the other hand,
in the large relative supercooling limit they differ even qualitatively:
$\RLcr$ vanishes, whereas $\RTcr$ goes to infinity. However, this limit is
beyond the validity of the expression (\ref{F},\ref{F2}) for the free energy,
in which all the temperature dependence comes from the pressure difference.

Both of the generalized surface tensions,
$\sigma _{{\rm L}} (R)$ and $\sigma _{{\rm T}} (R)$, are plotted
versus the {\em inverse} radius $l_c / R$ in \fig\ref{fig:sofr}.
For large values of the bubble radius the two surface tensions coincide.
In this limit, depending on the value of the constant $T_0 / T_c$, the surface
tensions can either decrease or increase with $R$, implying that the curvature
coefficient can be positive or negative, respectively (see \fig\ref{fig:gd}).
When the bubble  radius gets smaller,
the two surface tensions differ increasingly.
The curves for $\sigma _{{\rm T}} (R)$ turn backwards,
because with decreasing $\lbar$, $\RTcr$ when expressed in physical units
possesses a minimum, after which it begins to increase
(see \fig\ref{fig:rofl}).
For small values of the radius, $\sigma _{{\rm L}} (R)$ has to decrease
with $R$, in order to compensate the faster decrease of the term
$\Delta p(T) R^3$ in the free energy expansion in \eq(\ref{F2}).
And vice versa, $\sigma _{{\rm T}} (R)$ must increase
with decreasing $\lbar$ in the large relative supercooling regime.
The two-valuedness of $\sigma _{{\rm T}} (R)$ does not imply that
the distance of maximum tension would be a bad definition for
the bubble radius. The radius $R_{{\rm T}}$ agrees well with
the intuitive picture of the bubble, and furthermore,
as will be discussed later,
its minimum value can be used as an estimate
for the break-down distance of the free energy expansion.

The coefficients in the free energy expansion in \eq(\ref{F})
were determined from \eq(\ref{gddef}), employing both definitions
for the bubble radius. Within numerical accuracy, the results were equal.
The reason for this is that both of the radii, and both of the generalized
surface tensions as well, coincide in the limit of infinite bubble radius.
Below, results for the coefficients are presented only for the case where
the radius is defined by Laplace's relation.

We write $\sigma _{{\rm L}} (\RLcr (T))$ and $l_c / \RLcr (T)$
as Laurent expansions of $\lbar$ around $\lbar \! = \! 1$,
which corresponds to infinite radius.
We find that the coefficient of the curvature term is
\begin{eqnarray}
  \gamma & = &  \left[ \frac{2}{9} - \frac{f'(1)}{9} - \frac{u}{2}
       \right] \sigmainfty l_c
   \approx  \left( 1.49 \pm 0.01 - \frac{T_c^2}{2 T_0^2}
         \right) \sigmainfty l_c \; , \la{gamma}
\end{eqnarray}
where
\be
  u = \frac{ \alpha^2 }{ 9 \lambda \gamma - 2 \alpha^2 }
    = \frac{1}{2} \left( \frac{T_c^2}{T_0^2} - 1 \right) \; .
\ee
In \eq(\ref{gamma}), $f'(1)$ denotes $\dd f(\lbar) / \dd \lbar$ at
$\lbar \! = \! 1$, and its numerical value is computed from \eq(\ref{fnum}).
For the coefficient of the constant term in \eq(\ref{F}) we find
\begin{eqnarray}
    \delta & = & \frac{1}{54} \left\{ f''(1) - 2f'(1) - 6
       + 2 u \left[ 5f'(1) - 1 \right]  + 27 u^2 \right\}
       \sigmainfty l_c^2 \la{delta}  \\
  & \approx & \left( 0.88 \pm 0.03 - 0.90 \frac{T_c^2}{T_0^2}
         + \frac{T_c^4}{8 T_0^4} \right)  \sigmainfty l_c^2 \; .
     \la{deltanum}
\end{eqnarray}
The coefficients $\gamma$ and $\delta$ are shown in \fig\ref{fig:gd}.
If the temperature $T_0$ is near $T_c$, the cofficient $\gamma$ is in units
of $\sigmainfty l_c$ close to unity, and $\delta$ almost vanishes.
If $T_0 / T_c$ is in the vicinity of $0.58$, the zero-point of $\gamma$,
the constant term dominates over the curvature term in the free energy
expansion
even for rather large bubbles. When $T_0 / T_c$ decreases from $ \sim 0.4$,
$\gamma$ decreases and $\delta$ increases rapidly.

The magnitudes of the different terms in the truncated free energy expansion
are compared in \fig\ref{fig:divr}.
{}From the figure we can conclude that if $T_0 / T_c > 0.5$, the truncated
expansion for the free energy of a spherical bubble in \eq(\ref{F}) breaks
down when the bubble radius is of the order of $2 l_c$. If $T_0 / T_c$
is smaller, the truncated free energy expansion is valid only for much
larger bubbles.
The distance where the truncated free energy expansion breaks down is for
different values of $T_0 / T_c$ roughly equal to
the smallest tension radius of the critical bubbles.

Finally, we will investigate if the low temperature phase bubbles can exist
at temperatures higher than $T_c$. Here the full expression for
the free energy, \eq(\ref{F2}), is used instead of the truncated one.
Now the volume term in this equation is positive.
Therefore the necessary condition for the existence
of stable low temperature phase bubbles is that the surface term
(i.e. the latter term)
not be a monotonically increasing function of the radius $R$.
The surface term is plotted in \fig\ref{fig:fofr} for two values of
$T_0 / T_c$, and it behaves qualitatively in the same manner for
all other values of $T_0 / T_c$. In spite of the two-valuedness of
$\sigma _{{\rm T}} (R)$ it is clear that the
behaviour of the surface term $4 \pi \sigma (R) R^2$ shows no indication
of the existence of low temperature bubbles above $T_c$. --- The expression
for the free energy in \eq(\ref{F2}) should not, however, be used for too small
values of the bubble radius, because qualitatively it does not describe
critical bubbles of large relative supercooling. The qualitative description
breaks down when the bubble radius is again of the order of the smallest
value of $\RTcr$.

The result that the magnitude of the generalized surface term grows with
the bubble size is very natural (even though $\sigma (R)$ is a decreasing
function of the bubble radius for large bubbles if $T_0/T_c \! > \! 0.58$,
the combination $\sigma (R) R^2$ is an increasing one).
However, in the MIT bag model calculation by Mardor and
Svetitsky~\cite{Mardor&Svetitsky} it was found that the free energy as
a function of the bubble radius has a minimum for $T \! > \! T_c$.
Clearly the conclusion must be that either the spherical MIT bag model,
or any order parameter model with a potential as the one in \eq(\ref{V}),
describes the quark--hadron phase transition qualitatively incorrectly.

To summarize,
we have discussed bubble free energy in
cosmological first order phase transitions
which can be described by the quartic potential.
Numerical function for the free energy of critical bubbles
with an $\pm 0.1 \%$ accuracy has been presented.
Two different definitions for the bubble radius have been studied:
Laplace's relation and the distance of maximum tension.
The corresponding curvature-dependent surface tensions
$\sigma _{\rm L} (R)$ and $\sigma _{\rm T} (R)$ have been calculated,
and expanded to second order in $1/R$ around infinite bubble radius.
It has been shown that for typical values of
the constant $T_0 /T_c$\ \ $\sigma (R)$ is
for large $R$ a decreasing function (positive $\gamma$),
but that in principle it can also be an increasing function
(negative $\gamma$).
No indications have been found
of the existence of low temperature phase bubbles above $T_c$.
The assumption that all the temperature dependence in the free energy comes
from the pressure difference between the two phases
made it possible to solve the general free energy $F(R,T)$
of spherical bubbles
from knowledge of the critical bubbles only.

\bigskip

{\em Acknowledgements:} This research was supported by the Academy of Finland.
The author wishes to thank K. Kajantie for the inspiration to this work as well
as for useful comments, and T. Ala-Nissil\"{a}, A. Laaksonen and M. Laine for
discussions.

\newpage



\begin{figure}
\caption[x]{ Non-divergent part of the critical action.
  Solid curve is $f(\lbar)$, as defined in \eq(\ref{fdef}),
  and dashed curve is $[ f(\lbar) / \lbar^{3/2} ] / 6$~.\la{fig:f} }
\end{figure}
\begin{figure}
\caption[x]{ Definition of $R_{{\rm T}}$.
  Solid curve is the gradient part of the action density
  of a critical ($\lbar \! = \! 0.9$)--bubble and dashed curve is
  the potential part, the first and the latter term in \eq(\ref{S}),
  respectively.
  The dimensionless radius is $R' = M R$, where $M$ stands for
  the bosonic mass in the potential, $M^2 = \gamma (T^2 - T_0^2)$.
  The free energy density is shown in units of
  $T M^3 \alpha / \lambda^{3/2}$ (note that since $\lbar$ is fixed,
  so is $T/T_c$). From the figure, we can read that
  for $\lbar = 0.9$, $\RTcr ' \approx 7.3 $~.\la{fig:s3dens} }
\end{figure}
\begin{figure}
\caption[x]{ Radius of the critical bubble for $T_0/T_c = 0.58$,
  according to the two definitions.
  Dashed curve is $\RLcr / l_c$, and solid curve
  $\RTcr / l_c$.\la{fig:rofl} }
\end{figure}
\begin{figure}
\caption[x]{ The generalized surface tension from both definitions,
  $\sigma _{{\rm L}} (R)$ and $\sigma _{{\rm T}} (R)$,
  plotted versus the inverse radius $l_c/R$.
  Dashed curves show $\sigma _{{\rm L}} (R) / \sigmainfty$,
  and solid curves $\sigma _{{\rm T}} (R) / \sigmainfty$.
  Upper curves are for $T_0/T_c \! = \! 0.99$, and lower curves for
  $T_0/T_c \! = \! 0.58$~.\la{fig:sofr} }
\end{figure}
\begin{figure}
\caption[x]{ The coefficients $\gamma$ and $\delta$ in the free energy
  expansion in \eq(\ref{F}), for different values of $T_0/T_c$.
  Dashed curve is $\gamma / (\sigmainfty l_c)$,
  coefficient of the curvature term,
  and solid curve is $\delta / (\sigmainfty l_c^2)$,
  coefficient of the constant term.
  The width of the curve for $\delta$ shows the uncertainty
  in \eq(\ref{deltanum}),
  due to numerical determination of $f''(1)$
  (and to much less extent, of $f'(1)$).\la{fig:gd} }
\end{figure}
\begin{figure}
\caption[x]{ The distances at which magnitudes of the curvature term and
  the constant term equal the surface term
  in the truncated free energy expansion of \eq(\ref{F}),
  plotted versus $T_0/T_c$.
  Dashed curve shows the distance $R_{\gamma} / l_c$ at which
  $4 \pi \sigmainfty R_{\gamma}^2 \! = \! 8 \pi | \gamma | R_{\gamma}$,
  and solid curve the distance $R_{\delta} / l_c$ at which
  $4 \pi \sigmainfty R_{\delta}^2 \! = \! 16 \pi |\delta| $.\la{fig:divr} }
\end{figure}
\begin{figure}
\caption[x]{ Logarithm of the surface term in the free energy expansion
  in \eq(\ref{F2}),
  $\log [ 4 \pi \sigma (R) R^2 / (\sigmainfty l_c^2) ]$,
  as a function of the bubble radius $R / l_c$.
  In the same manner as in \fig\ref{fig:sofr},
  dashed curves represent $4 \pi \sigma _{{\rm L}} (R) R^2$,
  solid curves $4 \pi \sigma _{{\rm T}} (R) R^2$;
  upper curves are for $T_0/T_c \! = \! 0.99$, lower for
  $T_0/T_c \! = \! 0.58$~.\la{fig:fofr} }
\end{figure}

\end{document}